\documentstyle[epsfig]{mn}

\newif\ifAMStwofonts
\AMStwofontstrue

\def\be{\begin{equation}}
\def\ee{\end{equation}}

\def\etal{{\it et al.~}}

\def\gs{\mathrel{\raise1.16pt\hbox{$>$}\kern-7.0pt
\lower3.06pt\hbox{{$\scriptstyle \sim$}}}}
\def\ls{\mathrel{\raise1.16pt\hbox{$<$}\kern-7.0pt
\lower3.06pt\hbox{{$\scriptstyle \sim$}}}}
\def\gtsima{$\; \buildrel > \over \sim \;$}
\def\ltsima{$\; \buildrel < \over \sim \;$}
\def\prosima{$\; \buildrel \propto \over \sim \;$}
\def\gsim{\lower.5ex\hbox{\gtsima}}
\def\lsim{\lower.5ex\hbox{\ltsima}}
\def\simgt{\lower.5ex\hbox{\gtsima}}
\def\simlt{\lower.5ex\hbox{\ltsima}}
\def\simpr{\lower.5ex\hbox{\prosima}}

\def\pp{\noindent\parshape 2 0truecm 17truecm 2truecm 15truecm}
\def\rf#1;#2;#3;#4 {\par\pp#1, #2, #3, #4. \par}

\def\pr{\ref@jnl{Phys.Rev}}     

\def\ie{{\frenchspacing\it i.e. }}
\def\eg{{\frenchspacing\it e.g. }}

\def\href#1;#2 {{\bf #1} : {\em #2}}

\def\beq#1{\begin{equation}\label{#1}}
\def\eeq{\end{equation}}
\def\beqa#1{\begin{eqnarray}\label{#1}}
\def\eeqa{\end{eqnarray}}

\def\H2p{H$_2^+$ }

\def\mH2p{H_2^+}

\ifoldfss
  \ifCUPmtlplainloaded \else
    \NewTextAlphabet{textbfit} {cmbxti10} {}
    \NewTextAlphabet{textbfss} {cmssbx10} {}
    \NewMathAlphabet{mathbfit} {cmbxti10} {} % for math mode
    \NewMathAlphabet{mathbfss} {cmssbx10} {} %  "   "    "
  \fi
  \ifAMStwofonts
    \ifCUPmtlplainloaded \else
      \NewSymbolFont{upmath} {eurm10}
      \NewSymbolFont{AMSa} {msam10}
      \NewMathSymbol{\upi}     {0}{upmath}{19}
      \NewMathSymbol{\umu}     {0}{upmath}{16}
      \NewMathSymbol{\upartial}{0}{upmath}{40}
      \NewMathSymbol{\leqslant}{3}{AMSa}{36}
      \NewMathSymbol{\geqslant}{3}{AMSa}{3E}

       \let\le=\leqslant
       
    \fi
  \fi
\fi % End of OFSS

\ifnfssone
  \newmathalphabet{\mathit}
  \addtoversion{normal}{\mathit}{cmr}{m}{it}
  \addtoversion{bold}{\mathit}{cmr}{bx}{it}
  \newmathalphabet{\mathbfit} % math mode version of \textbfit{..}
  \addtoversion{normal}{\mathbfit}{cmr}{bx}{it}
  \addtoversion{bold}{\mathbfit}{cmr}{bx}{it}
  \newmathalphabet{\mathbfss} % math mode version of \textbfss{..}
  \addtoversion{normal}{\mathbfss}{cmss}{bx}{n}
  \addtoversion{bold}{\mathbfss}{cmss}{bx}{n}
  \ifAMStwofonts
    \ifCUPmtlplainloaded \else
      \UseAMStwoboldmath
      \makeatletter
      \new@mathgroup\upmath@group
      \define@mathgroup\mv@normal\upmath@group{eur}{m}{n}
      \define@mathgroup\mv@bold\upmath@group{eur}{b}{n}
      \edef\UPM{\hexnumber\upmath@group}
      \new@mathgroup\amsa@group
      \define@mathgroup\mv@normal\amsa@group{msa}{m}{n}
      \define@mathgroup\mv@bold\amsa@group{msa}{m}{n}
      \edef\AMSa{\hexnumber\amsa@group}
      \makeatother
      \mathchardef\upi="0\UPM19
      \mathchardef\umu="0\UPM16
      \mathchardef\upartial="0\UPM40
      \mathchardef\leqslant="3\AMSa36
      \mathchardef\geqslant="3\AMSa3E

       \let\le=\leqslant

    \fi
  \fi
\fi % End of NFSS release 1

\ifnfsstwo
  \DeclareMathAlphabet{\mathbfit}{OT1}{cmr}{bx}{it}
  \SetMathAlphabet\mathbfit{bold}{OT1}{cmr}{bx}{it}
  \DeclareMathAlphabet{\mathbfss}{OT1}{cmss}{bx}{n}
  \SetMathAlphabet\mathbfss{bold}{OT1}{cmss}{bx}{n}
  \ifAMStwofonts
    \ifCUPmtlplainloaded \else
      \DeclareSymbolFont{UPM}{U}{eur}{m}{n}
      \SetSymbolFont{UPM}{bold}{U}{eur}{b}{n}
      \DeclareSymbolFont{AMSa}{U}{msa}{m}{n}
      \DeclareMathSymbol{\upi}{0}{UPM}{"19}
      \DeclareMathSymbol{\umu}{0}{UPM}{"16}
      \DeclareMathSymbol{\upartial}{0}{UPM}{"40}
      \DeclareMathSymbol{\leqslant}{3}{AMSa}{"36}
      \DeclareMathSymbol{\geqslant}{3}{AMSa}{"3E}

       \let\le=\leqslant

    \fi
  \fi
\fi % End of NFSS release 2

\ifCUPmtlplainloaded \else
  \ifAMStwofonts \else % If no AMS fonts
    \def\upi{\pi}
    \def\umu{\mu}
    \def\upartial{\partial}
  \fi
\fi

\title[MIR Emission from the First Objects]{Detecting the First 
Objects in the Mid-IR with NGST}

\author[Benedetta Ciardi \& Andrea Ferrara]{
Benedetta Ciardi$^1$ and Andrea Ferrara$^2$\\
$^1$ Universit\'a di Firenze, Largo Enrico Fermi 5, 50125 Firenze, Italy \\
$^2$ Osservatorio Astrofisico di Arcetri, Largo Enrico Fermi 5, 50125
Firenze, Italy \\}

\date{November 2000}

\pagerange{\pageref{firstpage}--\pageref{lastpage}}
\pubyear{2000}
\begin{document}

\maketitle
\label{firstpage}
\begin{abstract}
We calculate the expected Mid-IR     molecular hydrogen line emission
from the first objects in the universe. Due to
their low masses, the stellar feedback from massive stars is able to 
blowaway their gas content and collect it into a cooling shell where
H$_2$ rapidly forms and IR roto-vibrational (as for example the restframe
2.12~$\mu$m) lines carry away a large fraction (up to 10\%)
of the explosion energy. The fluxes from these sources are in 
the range $10^{-21}-10^{-17}$~erg~s$^{-1}$~cm$^{-2}$.
The highest number counts are expected in the 20~$\mu$m
band, where about $10^5$ sources/deg$^2$ are predicted at the limiting flux
of $3 \times 10^{-18}$~erg~s$^{-1}$~cm$^{-2}$. 
Among the planned observational facilities, we find 
that the best detection perspectives are offered by NGST, 
which should be able to reveal
about 200 first objects in one hour observation time at its limiting flux 
in the above band. 
Therefore, Mid-IR         
instruments appear to represent perfect tools to trace star formation
and stellar feedback in the high ($z\simgt 5$) redshift universe.

\end{abstract}

\section{Introduction}

Detecting the first luminous objects in the universe will be the
primary goal of several future space- and ground-based telescopes. 
The importance of these experiments consists in the fact that they 
could be able to test current cosmological scenarios, study the
properties of these (supposedly very small)  galaxies, and their 
effects on the surrounding environment (as for example reionization,
heating and metal enrichment of the IGM). 
Due to the predicted low luminosity of such  objects, this task will 
be at the capability edge of even the most advanced and powerful instruments.  
Apart from the indirect probes of their effects, as for example the 
secondary anisotropies in the CMB left by reionization 
(Knox, Scoccimarro \& Dodelson 1998; Gruzinov \& Hu 1998; 
Bruscoli \etal 2000; Benson \etal 2000),
some search strategies have already been suggested in the literature.
Marri \& Ferrara (1998) and Marri, Ferrara \& Pozzetti (2000) have suggested 
that Type II supernovae occurring in the first objects might outshine their 
parent galaxy by more than a hundred times and become visible by
instruments like the {\it Next Generation Space Telescope} (NGST).
Schneider \etal (2000) have investigated the possibility of detecting
gravitational wave emission from high-redshift very massive objects with
LISA.
Oh (1999) proposes direct imaging of the ionized halos around primordial
objects either via their free-free emission (possibly detectable with 
the Square Kilometer Array)  or Balmer line emission again with NGST.

One point that is particularly important when dealing with these low
mass systems is that they are strongly affected by feedback mechanisms
both of radiative and stellar type; these have been extensively
investigated by Ciardi \etal (2000). Molecular hydrogen, being the only
available coolant in a plasma of nearly primordial composition, is
a key species in the feedback network as it regulates the collapse 
and star formation in these objects. Ferrara (1998, F98) pointed out 
that H$_2$ is efficiently
formed in cooling gas behind shocks produced during the blowaway (\ie
the complete ejection of the galactic gas due to SN explosions) process 
thought to occur in the first objects, with typical H$_2$ fractions $f_{H_2}
\approx 6 \times 10^{-3}$. We shall see that the conditions in these cooling blastwaves 
are such that a noticeable amount of the explosion energy is carried 
away by infrared (redshifted into the Mid-IR [MIR] spectral region) H$_2$ molecular lines, 
which therefore might provide us
with a superb tool to detect and trace these very distant primordial galactic 
blocks.

The use of molecular lines as diagnostics of moderate redshift ($z\simlt
3$) galaxies has already been proposed by some authors (Frayer \& Brown
1997; Blain \etal 2000); at higher redshifts pioneering calculations were
carried on by Shchekinov \& \'Entel' (1985); more recently Silk \& Spaans (1997)
concentrated on the CO and dust emission from HII regions inside larger
galaxies. All these studies have emphasized the power of the molecular
line emission as a probe of distant sources.
Based on previous calculations (F98), in this paper we calculate MIR        
fluxes and number counts for
these sources in various bands and assess if forthcoming observing
facilities will be able to eventually unveil the beginning of the 
cosmic star formation era     and study the feedback processes in the
young universe.

\section{Emission model}\label{emmod}

Due to its symmetry, the H$_2$ molecule has no electric dipole moment in the ground
state. Therefore, the first detectable H$_2$ emission lines 
are produced by quadrupole radiation and they are purely rotational.
The line properties of interest to the present work, both for rotational and
roto-vibrational lines, are listed in Table~1.
There we see that the excitation temperatures, $T_{ex}$, of these lines fall in the range 500-7000 K.
The temperature range spanned during the cooling of the post-shock IGM gas produced
during the blowaway of low-mass primordial galaxies is (F98):
\be
\label{ts}
300~{\rm K} \simlt T \simlt 2.3 M_6^{2/5} (1+z)^{18/5}~{\rm K}  
\ee
where $M=10^6 M_6~M_\odot$ is the total mass of the galaxy.\footnote{
We adopt a $\Lambda$CDM (cluster normalized) cosmology with 
$\Omega_M=0.35, \Omega_\Lambda=0.65, \Omega_b=0.04, h=0.65$.}
The upper limit is set by the condition that the cooling time of the
post-shock gas is shorter than the Hubble time.
Strictly speaking the above upper limit holds until the IGM cooling
is dominated by the inverse Compton radiative losses ($z\simgt 6$), but
we use it down to $z=4$, where it still represents a very good approximation for our purposes.
Given the range in eq.~\ref{ts}, it is then conceivable that the above molecular lines are 
excited during the process in which a cold shell forms, thus producing potentially detectable radiation.
The flux observed on the ground in a given line is derived as follows. 

Let $\nu_{ik}$ be the restframe frequency of photons emitted by the
molecule during the transition between the energy levels $k$ and $i$; then
the line emissivity (erg~s$^{-1}$~cm$^{-3}$~sr$^{-1}$) is (Spitzer 1978):
\be
j_{\nu_{ik}}=\frac{h}{4 \pi} \nu_{ik} n_k A_{ki},
\label{em}
\ee
where $n_k$ is the number density of molecules in the $k$ level and
$A_{ki}$ is the Einstein coefficient for spontaneous emission.
The typical H$_2$ densities found are much lower than the critical
one ($\approx 10^4$~cm$^{-3}$), hence we neglect collisional de-excitations.

As the conditions for thermodynamic equilibrium are not satisfied, the
population of the various levels must be obtained by solving the detailed
balance equations (Spitzer 1978). Using this approach, the number
density of molecules in the vibrational level $v$, $n_J$, of even and odd 
rotational levels decouple, and they are obtained
by iteration using the following formula:
\be
\label{boltz1}
n_{J+2}(v)= n_J(v)~ g_J~{\gamma_J {\rm e}^{-\Delta E/kT}\over 1+ \gamma_J},
\ee
where $g_J$ is the statistical weight of level $J$, $\gamma_J$ is the
ratio between the collisional excitation rate and the Einstein $A$
coefficient, $\Delta E$ is the energy difference between the two
levels. Obviously, in the limit of large $\gamma_J$ eq.~\ref{boltz1} 
approaches the Boltzmann distribution. 

Given the post-shock temperatures found (eq.~\ref{ts}), it is only necessary to consider
the two vibrational levels $v=0,1$. The relative number of molecules in
these two levels depends on the total molecular hydrogen density,
$n_{H_2}$.  For a given redshift, the value of $n_{H_2}$ (F98) is: 
\be
n_{H_2}(z)=p\;f_{H_2}(M,z) n_s \simeq p\;f_{H_2}(M,z) \delta~ n_H(z), 
\label{nh2}
\ee
where $f_{H_2}$ is the molecular fraction in the shell, and $n_H(z)$ is the IGM
hydrogen density. We allow for a density enhancement in the shell with respect 
to $n_H$ equal to $\delta$; this is produced both by the shock compression
and the possible occurrence of the explosion inside an overdense region of the 
universe (\eg a cosmological filament). Because of the first effect, and
as the expanding blastwave becomes rapidly radiative, $\delta$ is given
by the square of the shock Mach number, ${\cal M}$, with respect to the 
ambient IGM. Following F98, it is easy to show that the density
enhancement is given by: 
\be
\label{delta} 
\delta = 4.8 M_6^{2/5} (1+z)^{13/5}.
\ee
For simplicity, we have neglected the additional density increase produced
by an explosion in a overdense region; thus, the results presented above
should be interpreted as a conservative lower limit to the detectability
of the sources under investigation.
Finally,  we assume a branching ratio $p=0.75$ (0.25) for ortho (para) 
transitions which is valid under LTE conditions; however, non-equilibrium
conditions might lead to slightly lower values of the ortho-to-para
ratio (Chrysostomou \etal 1993; Rodr\'iguez-Fern\'andez \etal 2000). 

\begin{table}
\label{tab1}
\centerline{Table 1: Relevant H$_2$ Transition Lines}
\begin{center}
\begin{tabular}{lccc}
\hline
\hline
Transition & $\Lambda^{(1)}$ [$\mu$m] & T$_{ex}^{(2)}$ [K] & A$^{(3)}$ [s$^{-1}$] \\
\hline
\hline
0-0S(0) & 28.0 &  512 & 2.94 $\times 10^{-11}$   \\
0-0S(1) & 17.0 & 1015 & 4.76 $\times 10^{-10}$   \\
0-0S(3) & 9.7  & 2503 & 9.84 $\times 10^{-9}$   \\
0-0S(5) & 6.9  & 4586 & 5.88 $\times 10^{-8}$   \\ 
1-0S(1) & 2.12 & 6953 & 3.47 $\times 10^{-7}$   \\ \hline
\end{tabular}
\end{center}
$^{(1)}${Emission wavelength}\\
$^{(2)}${Excitation temperature (Combes \& Pfenninger 1998;
Timmermann \etal 1996)}\\
$^{(3)}${De-excitation Einsten coefficient (Turner, Kirby-Docken
\& Dalgarno 1977)}\\
\end{table}
With the emissivity given by eq.~\ref{em}, the observed flux (erg~s$^{-1}$~cm$^{-2}$) 
is:
\be
F(\nu_o)=\frac{L_{\nu_{ik}}(1+z)}{d_L^2 },
\label{flux}
\ee
where $\nu_o=\nu_{ik}/(1+z)$ is the observed frequency, $L_{\nu_{ik}}$ the
luminosity and $d_L$ is the cosmological luminosity
distance. For a flat universe ($\Omega_0=\Omega_M+\Omega_\Lambda=1$),
$d_L$ is given by
\be
d_L=(1+z)\int_0^z (1+z^\prime)\left \vert {dt\over dz^\prime}\right \vert dz^\prime,
\label{delle}
\ee
\be
\left \vert {dt\over dz}\right\vert^{-1} = H_0(1+z)\sqrt{(1+\Omega_Mz)(1+z)^2-\Omega_\Lambda z(2+z)},
\label{delle1}
\ee
where $H_0=100h$ km s$^{-1}$ Mpc$^{-1}$ is the current Hubble constant.
The corresponding luminosity is:
\be
L_{\nu_{ik}}=4 \pi V j_{\nu_{ik}},
\label{lum}
\ee
where $V=(4/3)\pi R_s^2 dR_s$ is the physical volume occupied by the H$_2$
forming shell.  The values of $R_s$ and $dR_s$ are obtained from the
formulae in F98;
the only difference consists in the assumption of a single burst of star
formation, as opposed to the quiescent star formation prescription 
of F98. By assuming a star formation efficiency of 10\%, and one  SN every 100
$M_\odot$ of stars formed, it is easy to show that the total energy of the explosion
is equal to $E=1.3\times 10^{53} M_6$~erg. Then, from F98 it follows:
\be
R_s \simeq 0.88 M_6^{1/5} (1+z)^{-11/5} \; {\rm Mpc},
\label{r}
\ee
\be
dR_s=\frac{N_{H_2}}{n_{H_2}} \simeq {f_{H_2} M_s \over m_H R^2_s n_{H_2}}, 
%={4\pi\over 3\delta p} R_s,
\label{dr}
\ee
where we have assumed that all the swept IGM mass is in the cool shell
of total mass $M_s$, together with the galaxy interstellar medium.
This result is somewhat dependent on the assumption made to derive the
energy of the explosion, while it is not strongly dependent on the
cosmology.

\section{Number Counts} \label{numbc}

The number of objects whose observed flux is larger than the threshold
value $F_{min}$ is:
\begin{eqnarray}
N(>F_{min}) & = & \int^{z_{max}}_{z_{min}} dz {dV \over dz} \int^{L_{max}}_{L_{min}}
dL {dn \over dM} {dM \over dL}  \nonumber \\
 & = & \int^{z_{max}}_{z_{min}} dz {dV \over dz} \int^{M_{max}}_{M_{min}}dM
 {dn \over dM},
\label{nc}
\end{eqnarray}
where $dn/dM$ is the comoving number density of halos with masses between $M$
and $M+dM$ (Press \& Schechter 1974) and $dV/dz$ is the
comoving volume element per unit redshift:
\be
{dV \over dz}= {4 \pi c d_L^2 \over (1+z)} \left \vert {dt \over dz} \right
\vert,
\label{vol}
\ee
where $d_L$ and $dt/dz$ are given in eqs.~\ref{delle} and~\ref{delle1},
respectively. The limits $z_{min}$ and $z_{max}$ depend on the
observational wavelength band; $L_{max}$ is the luminosity
corresponding to the maximum mass value, $M_{max}$, 
experiencing  blowaway  at redshift $z$ (Ciardi \etal 2000); $M_{min}={\rm
min}[M_H,M(L_{min})]$. Here $M_H$ is the minimum halo mass in which stars can
form at a given redshift, which is related to the minimum virial temperature of
$\approx 10^4$~K below which atomic hydrogen cooling of the gas is suppressed
and fragmentation into stars is inhibited (Haiman, Rees \& Loeb 1997; Ciardi
\etal 2000); $M(L_{min})$ is the mass of the halo producing the
luminosity $L_{min}$ corresponding to the observed flux $F_{min}$.
Finally, the luminosity, $L$, is taken    as the maximum one reached by
the object (usually corresponding to temperatures equal to the excitation
ones) that experiences a blowaway at redshift $z$, providing that it is 
emitted at a redshift $> z_{min}$. These calculations give
lower limits, as they do not include objects blowing away at
$z>z_{max}$, but having a peak luminosity at $z<z_{max}$.

\section{Results}\label{res}

\begin{figure}
\vskip -2.truecm
\psfig{figure=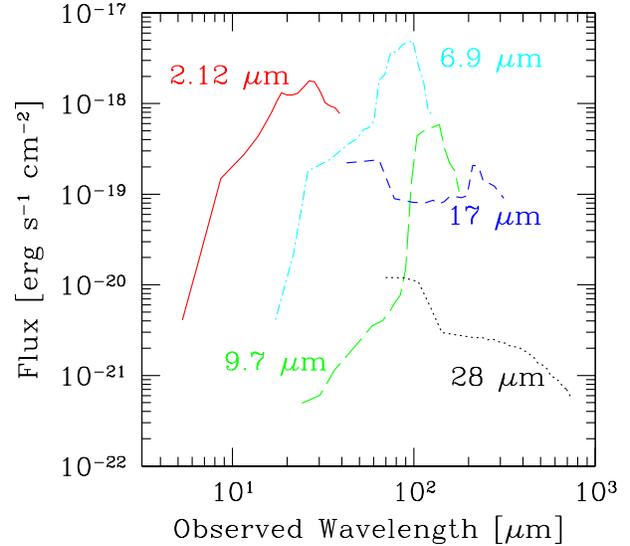,height=10cm}
\caption{\label{fig1}\footnotesize{
Expected flux as a function of the observed wavelength for the five
H$_2$ molecular lines considered: $\lambda_e=2.12$~$\mu$m (solid  line),
6.9 $\mu$m (dot-dashed), 9.7 $\mu$m (long dashed), 17 $\mu$m (short-dashed)
and 28 $\mu$m (dotted).
Fluxes are shown for halos with mass equal to the mean one at the 
relevant emission redshift (see text for details). 
}}
\end{figure}
The above model allows us to determine the observational
properties of the MIR        emission from the first objects. In 
particular, we present results concerning the expected flux in the
various lines considered (see Table~1) and the corresponding number
counts. These are then compared with the sensitivities of various
planned instruments. 

\subsection{Expected MIR        Fluxes}

Fig.~\ref{fig1} shows the expected flux as a function of the observed 
wavelength, $\lambda_{o}$, in the five different lines 2.12 $\mu$m, 
6.9 $\mu$m, 9.7 $\mu$m, 17 $\mu$m and 28 $\mu$m.
As an example, we plot the fluxes calculated for halos of mass $\approx 10^8-10^9 M_\odot$,
corresponding to the mean one among those suffering blowaway,
as given by the Press-Schechter formalism, at the explosion
redshift. 
The curves refer to emission occurring when 
the shocked gas temperature has decreased down to values close 
to the corresponding line
excitation temperature. Thus, the actual                emission
redshift is somewhat lower than the explosion redshift.

From Fig.~\ref{fig1} we see that the typical observed fluxes are in the range
$10^{-22}-10^{-17}$ erg~s$^{-1}$~cm$^{-2}$, 
depending on the wavelength band (and therefore emission redshift). 
For the 2.12 $\mu$m, 6.9 $\mu$m and 9.7 $\mu$m emission 
lines the flux is increasing with
wavelength, whereas the flux of the 
17 $\mu$m and 28 $\mu$m
excitation lines is decreasing towards higher $\lambda_{o}$.
From a general point of view this can be understood as follows. As
the IGM density increases with redshift, this causes a corresponding
increase in the emissivity. 
In addition, the postshock gas tends to be warmer at high $z$, hence
inducing preferentially the excitation of the higher excitation
temperature lines. These effects are dominant for the 2.12 $\mu$m, 6.9
$\mu$m and 9.7 $\mu$m lines, while they are overwhelmed by a larger value
of $d_L$ and a smaller shell emission volume for the redder ones, producing a 
decreasing intensity. 

\begin{figure}
\vskip -2truecm
\psfig{figure=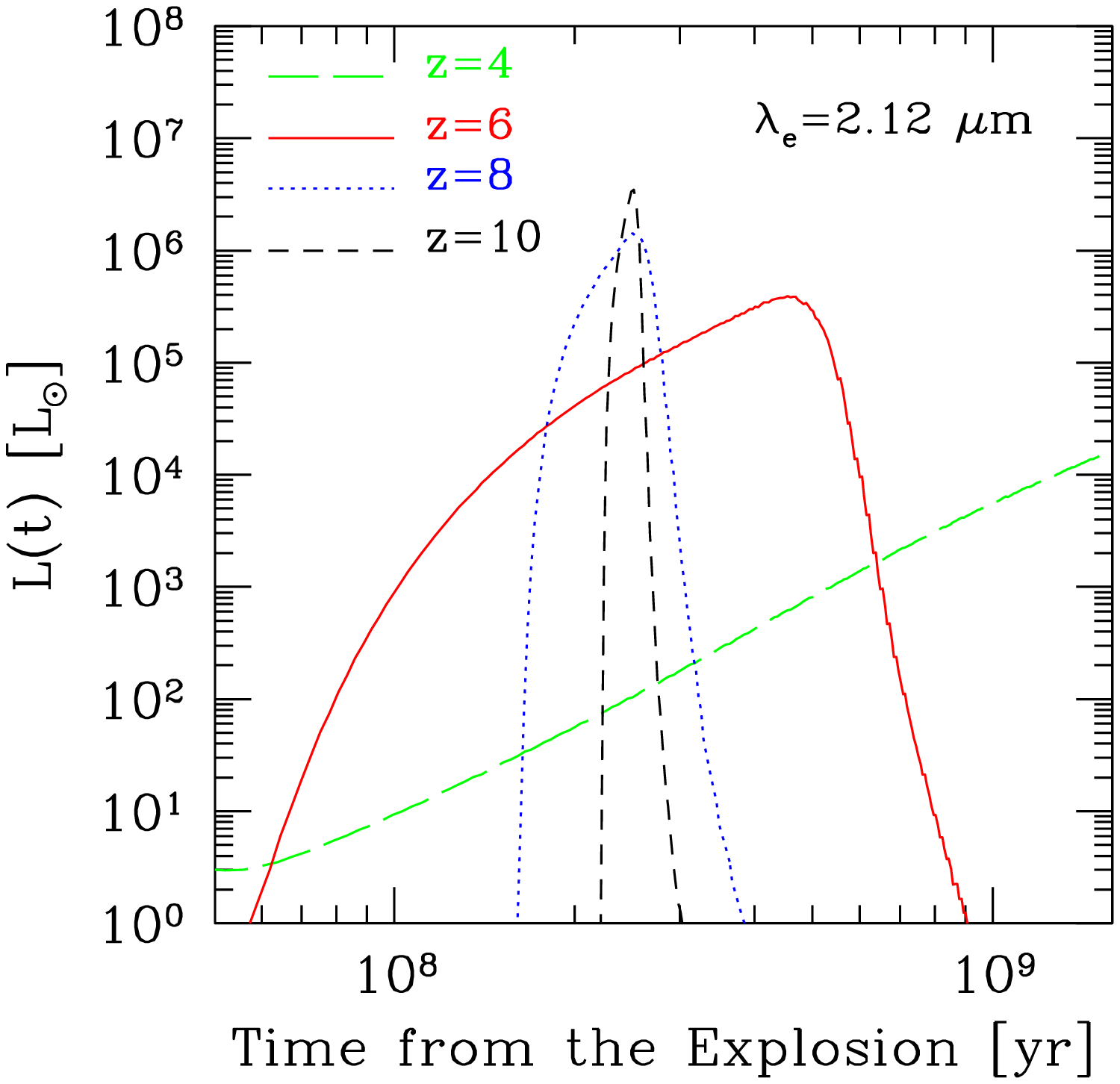,height=10cm}
\vskip -2truecm
\psfig{figure=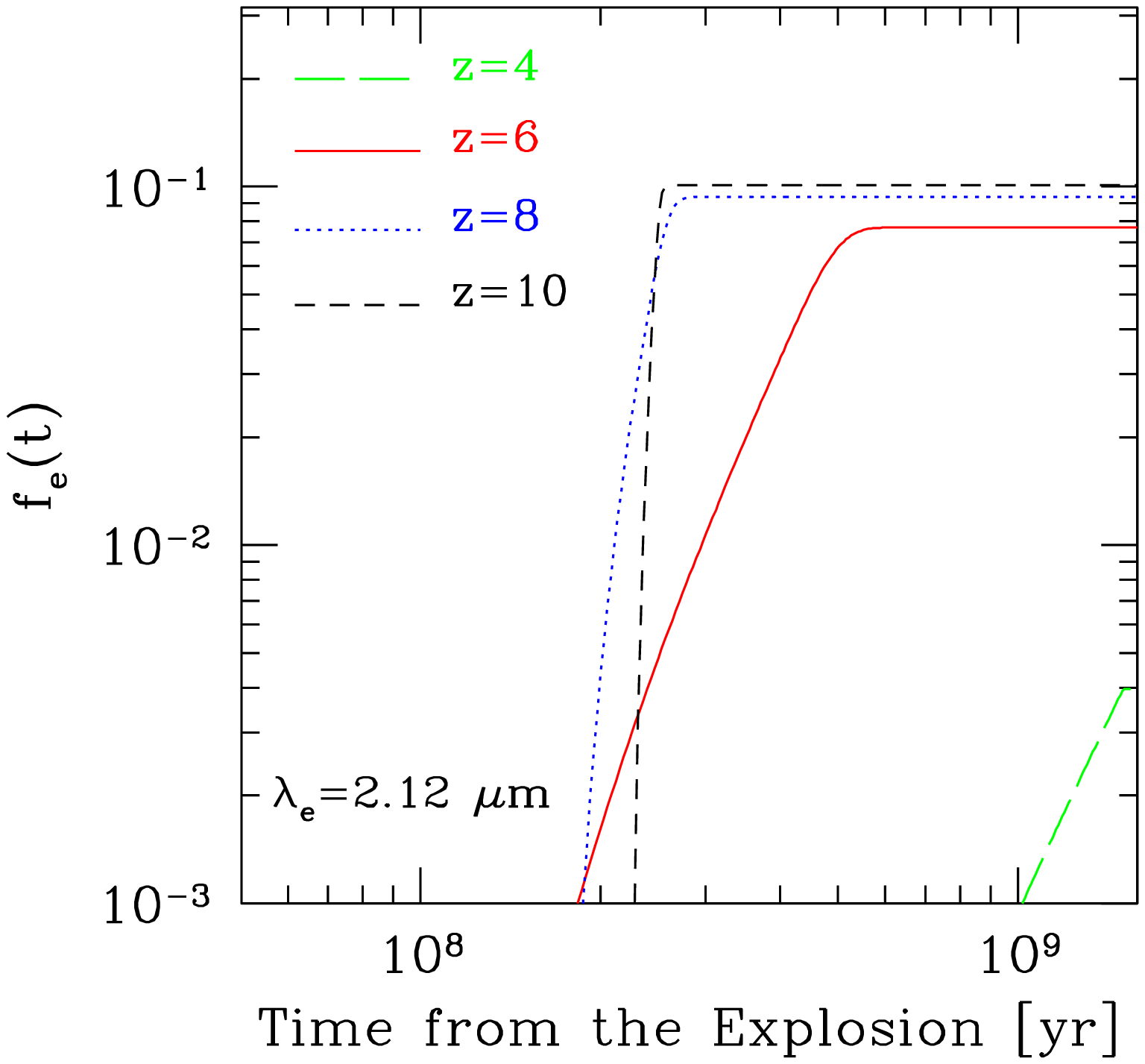,height=10cm}
\caption{\label{fig2}\footnotesize{(a) Luminosity evolution 
for a blown-away object at $z=4$ (long-dashed line), $z=6$ (solid), $z=8$ (dotted) and 
$z=10$
(short-dashed). The emission wavelength is $\lambda_{e}=2.12$~$\mu$m and
the luminosity is evaluated for the average halo mass (see text). 
(b) Fraction of the SN mechanical energy emitted in the $\lambda_e=2.12$~$\mu$m line
as a function of time. Notation is the same as (a). }}
\end{figure}
As detection chances are highest   for the 2.12~$\mu$m line, we have
derived its luminosity evolution and the corresponding fraction of the
SN energy radiated in this line by objects at different 
redshifts. As for Fig.~\ref{fig1}, the emission is derived for an average
halo mass. 
In Fig.~\ref{fig2}a we show the luminosity evolution for selected redshifts in
the range $4 \le z \le 10$. 
As the explosion reshift is increased, the peak of the emission is shifted at
earlier times and it becomes narrower. Both effects are due to the faster
evolution of the temperature driven by the enhanced (\ie $\propto (1+z)^4$) Compton 
cooling rates.
The more pronounced peak occurring at higher $z$ is due to the  $(1+z)^3$ density increase;
this trend    is partially counterbalanced by the decreasing average halo
masses with redshift. 
Fig.~\ref{fig2}b shows the time dependence of the cumulative fraction 
of the SN mechanical energy 
 $f_{e}(t)=E_{e}(t)/E$ emitted in the $\lambda_e=2.12$~$\mu$m line.
This quantity increases over a time
scale inversely related to the explosion redshift and reaches a plateau, determined
by the position of the $L(t)$ peak seen in Fig.~\ref{fig2}a. 
At high redshift up to 10\% of the SN mechanical energy is carried away by the 
considered roto-vibrational line; at lower redshifts this value rapidly decreases
below 1\%.

\subsection{Number Counts and Detectability}

\begin{figure}
\vskip -2.truecm
\psfig{figure=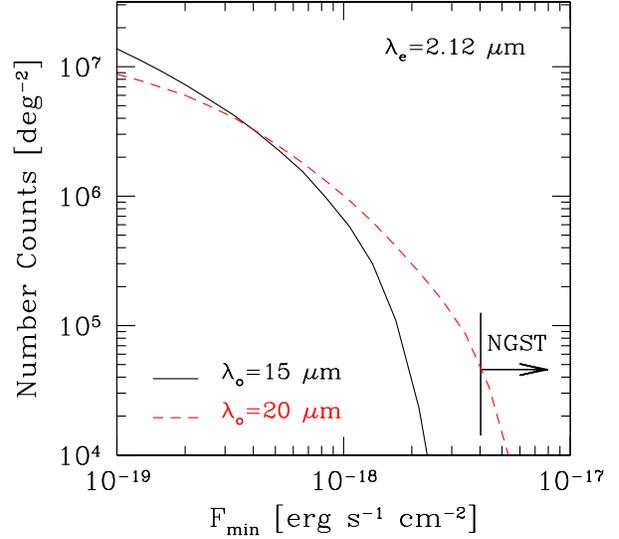,height=10cm}
\caption{\label{fig3}\footnotesize{Number counts from the H$_2$ 
$\lambda_e=2.12$~$\mu$m
emission wavelength as a function of the minimum observed flux, $F_{min}$.
The number counts are calculated for different observed wavelength bands
with $R=3$ and centered at $\lambda_{o}=15$~$\mu$m (solid line) and 20
$\mu$m (dashed). The limiting sensitivity of NGST in the same bands is
also shown.}}
\end{figure}
By using the expressions given in eq.~\ref{nc} we calculate
the expected number of objects per unit sky area as a function of the 
limiting flux, $F_{min}$, of a given experiment. 
The corresponding curves are shown in Fig.~\ref{fig3} for the 2.12~$\mu$m H$_2$ line. 
The number counts have been estimated in two  different
MIR      bands  in               the sensitivity range of planned imaging detectors
on board of the 
{\it Next Generation Space Telescope}\footnote{\tt http://augusta.stsci.edu}
(NGST). 
For our study, we consider the $R=3$ imaging mode of NGST, in the
two bands centered at $\lambda_{o}=15$ and 20 $\mu$m. 
In the $20~\mu$m band, the largest fluxes are at the
$6 \times 10^{-18}$ erg~s$^{-1}$~cm$^{-2}$ level, where 
about $10^4$ objects/deg$^2$ should be seen;   
the surface density of sources  increases rapidly as $F_{min}$ decreases.
In  the $15~\mu$m band objects are fainter and a rapid drop
is seen in the number counts above  $10^{-18}$ erg~s$^{-1}$~cm$^{-2}$.

In the above bands, NGST will           reach a sensitivity of $\simeq 4
\times 10^{-18}$~erg~s$^{-1}$~cm$^{-2}$ (shown in the Figure) for a $3\sigma$ detection in one
hour.  At the above sensitivity level, and with its field of view
of 4'$\times$4', NGST should be able to detect 
about 200 sources in the $20 \mu$m band
in one hour integration at the $3\sigma$ level. 
These objects are all located
in the redshift range $\Delta z \simeq 7-10$, a very intriguing time during
cosmic evolution, perhaps bracketing the reionization epoch. 
Assuming a diffraction limited telescope (angular resolution $\approx 0.25"$)
at this wavelength, field crowding should not be a problem, the number of sources being
well below the confusion limit.  

Longer exposure times might also allow detections in the 15 $\mu$m band. 
Sensitivities comparable to the NGST ones are also expected
for the SPICA HII/L2 future mission\footnote{\tt
http://www.ir.isas.ac.jp.}. Other MIR planned facilities, such as
SIRTF or FIRST will only reach sensitivities $\simeq 10^{-14}-10^{-15}$
erg~s$^{-1}$~cm$^{-2}$, depending on observational bands.

In conclusion, NGST appears to constitute the best tool to reveal
the emission from the first objects; while other  above mentioned instruments  
will not reach the required sensitivities.                      

\section{Conclusions}

We have investigated the possibility to detect the first objects in the
universe through the MIR line emission associated with the  
stellar feedback (\ie blowaway of the gas) induced by their supernovae.
A noticeable fraction of the explosion energy (up to 10\% in the 2.2$~\mu$m line alone) 
is eventually converted
and carried away by molecular hydrogen 
roto-vibrational lines. By using a detailed treatment of the 
non-equilibrium formation and emission of H$_2$ molecules behind these
cosmological blastwaves we have calculated the observed fluxes and
number counts of primordial objects and compared them with the expected
sensitivities of future instruments.             
At the limiting flux of NGST, we do predict that 
about 200
primordial objects can be detected 20~$\mu$m
in barely one hour of observation time. 

This detection would allow to directly image the first star formation
in the universe and our results show that MIR observations, when
compared with estimates of previous studies in other bands and/or
exploiting different strategies, represent a superb tool for this study. 
In addition, the proposed excitation mechanism will also allow to
test and calibrate the stellar feedback process in the dark ages and
finally assess the degree at which the first galaxies (and the IGM) have been
influenced by supernova energy deposition. This obviously holds the key
for the understanding of the subsequent evolution and formation of 
larger galactic blocks in most cosmological models. 
MIR        line emission allows the high redshift universe ($z\simgt
5$) to be much more easily explored than in the near IR bands, 
where the intergalactic absorption might be found to strongly affect the
detectability of objects located close or beyond the reionization epoch:  
for example, the predicted dust opacity to sources located at redshift $\sim 5$
is as high as $\sim 0.13$ at the observed wavelength 
$\lambda_0 \sim 1 \mu$m, and could considerably
affect observations of the distant universe in that band (Ferrara \etal 1999; 
Loeb \& Haiman 1997).
It is worth noting that our predictions constitute lower limits to the
number of observable objects as we have not considered possible density
enhancements descending from explosions occurring inside overdense
regions, as the filaments of the cosmic web (and also because of some 
details of the calculations, see Sec. 3).

Molecular hydrogen lines might also be excited during the collapse of
the so-called PopIII objects, which rely on this molecule to collapse
and form stars. However, there are several reasons to
suspect that the contribution of this process will be negligible with
respect to the emission due to the stellar feedback. First, if a
UV background is present, molecular hydrogen can be destroyed by
photodissociation. Even if there is no UV background, dissociation
of molecular hydrogen by internal UV radiation emitted from massive stars,
formed in the high density regions of the objects, is
very efficient and the evolution of the objects and star formation are
strongly affected (Omukai \& Nishi 1999). This regulation effect is efficient
as long as the line emission of molecular hydrogen is the main cooling
process, \ie if $Z \lsim 10^{-2} Z_\odot$ and $T \lsim 8000$ K
(Nishi \& Tashiro 2000). Second, the mass of these objects is very small
and therefore the amount of gas in the required thermodynamic state
for the emission is very limited. Interestingly, the relative little
importance of PopIII objects in terms of their radiation power in
ionizing photons with respect to objects of mass above $M_H$ has already
been established by Ciardi \etal (2000).  The latter objects emit a very
large fraction of their binding energy in the hydrogen Ly$\alpha$ line
which is therefore a much better tracer of their formation 
(Haiman, Spaans \& Quataert 2000).

\section*{Acknowledgments}

We are indebted to the referee, R. Schneider, for insightful comments.
It is also a pleasure to acknowledge several useful discussions with A. Loeb, 
R. Maiolino, R. Nishi, F. Palla, N. Scoville, Y. Shchekinov and T. Takeuchi. 
AF acknowledges support from Ecole Normale Superieure, Paris.

\newpage

\end{document}